\documentclass[reprint, amsmath, amssymb, prl]{revtex4-1}
\usepackage{amsfonts,amsthm}
\usepackage[svgnames]{xcolor}        
\usepackage[utf8]{inputenc}          
\usepackage[parfill]{parskip}        
\usepackage{graphicx}                
\usepackage{float}                   

\usepackage{hyperref}
\hypersetup{
    pdftitle={Electron-Beam Driven Relaxation Oscillations in Ferroelectric Nanodisks},
    pdfauthor={Nathaniel Ng, Rajeev Ahluwalia, Ashok Kumar, David J. Srolovitz, Premala Chandra and James F. Scott},
    pdfkeywords={nanodomain reorientation} {relaxation oscillations} {faceting} {ferroelectric nanodisks},
    linkcolor=red,
    urlcolor=blue,
    citecolor=brown,
    colorlinks=true
}
\begin{document}
\title{Electron-Beam Driven Relaxation Oscillations in Ferroelectric Nanodisks}
\author{Nathaniel Ng}
\author{Rajeev Ahluwalia}
\affiliation{Institute of High Performance Computing, Singapore 138632, Singapore}
\author{Ashok Kumar}
\affiliation{CSIR-National Physical Laboratory, Delhi India}
\author{David J. Srolovitz}
\affiliation{Departments of Materials Science and Engineering, Mechanical Engineering and Applied Mechanics, University of Pennsylvania, Philadelphia, Pennsylvania 19104, USA}

\author{Premala Chandra}
\affiliation{Center for Materials Theory, Department of Physics and Astronomy, Rutgers University, Piscataway, NJ 08854 USA}

\author{James F. Scott}
\affiliation{Department of Physics, Cavendish Laboratory, J. J. Thompson Avenue, Cambridge CB3 0HE, United Kingdom}
\affiliation{Departments of Chemistry and Physics, University of St. Andrews, St. Andrews YX16 9ST, United Kingdom}
\date{\today}

\begin{abstract}
Using a combination of computational simulations, atomic-scale resolution imaging and phenomenological modelling, we examine the underlying mechanism for nanodomain restructuring in lead zirconate titanate (PZT) nanodisks driven by electron beams.  The observed subhertz nanodomain dynamics are identified with relaxation oscillations where the charging/discharging cycle time is determined by saturation of charge traps and nanodomain wall creep. These results are unusual in that they indicate very slow athermal dynamics in nanoscale systems.
\end{abstract}
\maketitle
\begin{figure}[htdp]
  \vspace{0pt}
  \begin{center}
    \includegraphics[width=\columnwidth]{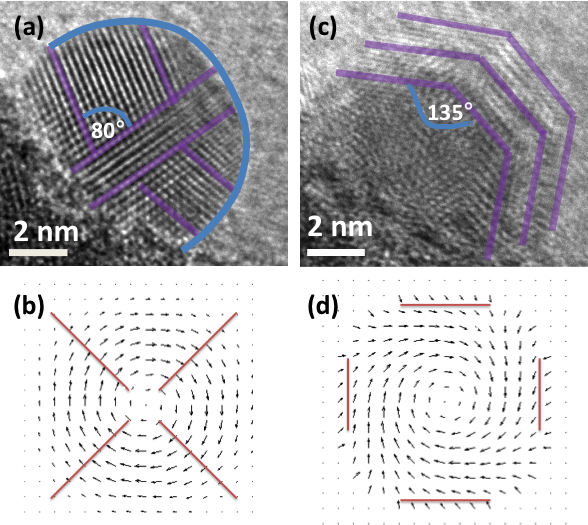}
  \end{center}
  \vspace{0pt}
  \caption{HRTEM images for (a) circular (unfaceted) and
  (c) faceted 8-nm PZT dots; the ferroelastic domain walls are illustrated as purple lines, and the angles between them are indicated.
  Results from simulations (distribution of polarization vectors) for (b) uncharged and (d) charged nanodisks; red lines indicate the approximate positions of domain walls.
  Domain walls are \textit{perpendicular} to the surface in the \textit{unfaceted} cases (a. and b.), but \textit{parallel} to the surface in the \textit{faceted} cases (c. and d.)}
\label{fig:hrtem-phasefield}
\end{figure}

The importance of finite-size boundaries for static polarization configurations has been emphasized in flux-closure \cite{Naumov04,Aguado-pente08,Jia11,McGilly11,Tang15} and faceted \cite{Ganpule02,scott2008,Lukyanchuk14} domain pattern studies in several ferroelectric materials.  Recently it has become possible to reorient spatially domains and defects on nanoscales with controlled high resolution transmission electron microscopy (HRTEM) \cite{robertson,Ahluwalia2013PRL,kumar2014}.  Nanodomain dynamics in ferroelectric nanodisks can also be probed with real-time HRTEM, and have been observed to be considerably slower (10 sec) \cite{kumar2014} than those in their nanomagnetic counterparts (50 nanoseconds)\cite{Bisig13}. Although in principle, such slow nanodomain reorientation could arise from thermal effects, it has been shown that such heating is negligible for our experimental conditions \cite{zheng}.
Here we emphasize that the observed driven domain dynamics of such free-standing ferroelectric nanostructures are strongly influenced by electrostatic conditions. In a combined computational, experimental and phenomenological effort, we identify the charging/discharging mechanism underlying the observed oscillations between circular and hexagonal nanodomain patterns in lead zirconate titanate (PZT) nanodisks driven by electron beams. Using phase-field modelling \cite{Ahluwalia2013PRL,Sriram2011ACSNano,Ng2012Acta}, we demonstrate the crucial influence of charge on the nanodomain patterns, and show that we can reproduce domain structures similar to those observed in experiment. The subhertz frequency response of the nanodomain reorientation results from relaxation oscillations \cite{Losev25,vanderPol,Pippard79} between threshhold charging of surface traps \cite{Li06} and subsequent discharging by nanodomain wall creep \cite{paruch}, where quantitative agreement is made using prior measurements on PZT samples.
Experimentally edge-supported 8-nm diameter disks of PZT were exposed to constant HRTEM beams with a $0.5\;\text{A/m}^2$ probe current density where further detail can be found elsewhere \cite{kumar2014}.  Two distinct nanodomain patterns were observed: the nanodomain walls ``flop'' between being normal to the disk perimeter to being parallel to it as displayed in Figure \ref{fig:hrtem-phasefield}a and \ref{fig:hrtem-phasefield}c.  Since HRTEM only reveals ferroelastic domain walls, modelling is needed to examine the full charging role played by the electron beam that drives the observed faceting oscillations and the simultaneous realignment of nanodomain walls.  It is known that electron beams can lead to charging of insulating materials \cite{Sessler2004}. Furthermore in the context of ferroelectrics, it has been shown that electron beams can be used to switch the polarization direction \cite{bonnell} and can even stabilize unusual quadrant patterns in free standing nanodisks \cite{Ahluwalia2013PRL}.  In order to investigate e-beam induced charging effects in ferroelectric nanodisks, we use a phase-field approach to simulate how domain patterns are influenced by radial fields generated by uniform free charge. The model, based on earlier studies \cite{Ahluwalia2009Nano, Ng2012Acta, Ahluwalia2013PRL}, has been adapted to incorporate the circular geometry (see supplementary material \cite{supplementary1}) of the experiment; the goal is to determine whether the observed nanodomain pattern realignment is driven by depolarization fields in a charging/discharging cycle.
We first simulate the uncharged case by solving the phase field equations for a system quenched from the paraelectric state.  A uniform charge density, $N_e = 4 \times 10^{26}\text{ m}^{-3}$, is then applied and the evolution of the domains is monitored. Fig \ref{fig:hrtem-phasefield}(b) shows the simulated pattern for the uncharged case with a flux closure domain pattern where the solid red lines indicate the positions of the ferroelastic domain walls, normal to the disk perimeter.  We note the excellent agreement with the domain wall configuration in the HRTEM image in Fig. \ref{fig:hrtem-phasefield}(a), suggesting that it corresponds to an uncharged disk. Figure \ref{fig:hrtem-phasefield}(d) shows the nanodomain pattern when the nanocrystal is charged ($N_e = 4 \times 10^{26}\text{ m}^{-3}$), i.e. during the charging process, the flux closure domain pattern transforms to one where polarization vectors have a non-zero radial component at the perimeter and a polarization vortex type state, characterized by a vanishing radial component, in the center.
This may be understood in terms of the radial electric field that vanishes at the center.  This radial field arises due to the uniformly charge density \cite{Ahluwalia2013PRL} associated with the electron beam induced charging. Vortex formation at the center occurs to avoid head-to-head polarization, resulting in areas having $\mathbf{P \cdot n} = 0$ (center) and $\mathbf{P \cdot n} \ne 0$ along the perimeter. Here, a radial unit vector is introduced as $\mathbf{n}=\frac{\mathbf{r}-\mathbf{r_0}}{\left|\mathbf{r}-\mathbf{r_0}\right|}$, where $\mathbf{r_0}=\left(x_0,y_0\right)$ represents the coordinates of the centre of the disc. Regions with such distinct $\mathbf{P \cdot n}$ are separated by ferroelastic nanodomain walls that are accessible to the electron beam technique. The solid red lines in Fig \ref{fig:hrtem-phasefield}(d) indicate such ferroeleastic domain walls which are aligned along the perimeter of the nanocrystal. Thus this situation corresponds to the HRTEM image in \ref{fig:hrtem-phasefield}(c), indicating that it results from nanodisks with charged boundaries.

\begin{figure}[htdp]
  \vspace{0pt}
  \begin{center}
    \includegraphics[width=\columnwidth]{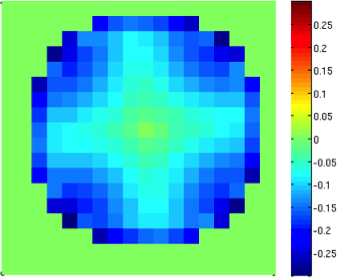}
  \end{center}
  \vspace{0pt}
  \caption{Distribution of the normalized radial component
  of the polarization,
  $\frac{1}{P_s}\mathbf{P \cdot n}$ corresponding to Fig 1d. The radial component of the polarization is zero in the center and nonzero towards the edges.  Note that the distribution shows strong anisotropy (i.e. $\theta$-dependence)}
\label{fig:radial-polarizations}
\end{figure}

While this charging/discharging scenario describes the realignment of domain walls, it does not directly model the observed kinetic faceting.  We emphasize its dynamical nature since two-dimensional faceting cannot occur in thermal equilibrium at finite temperatures \cite{berge}. However, in Fig \ref{fig:hrtem-phasefield}(d), we see that the distribution of $\mathbf{P \cdot n}$ is anisotropic, namely strongly $\theta$-dependent, and this is further emphasized graphically in Figure \ref{fig:radial-polarizations}.

Indeed the distribution of elastic energy densities will also be orientation-dependent due to the underlying crystal axes of PZT. More generally kinetic faceting is dynamically selected to accommodate external anisotropies that could also include applied stress and substrate-related issues \cite{Ganpule02,scott2008,Lukyanchuk14} where details depend on the specifics of the experimental situation and the materials involved; indeed faceting nanodomain patterns were not observed in HRTEM experiments on PZT nanotargets with square and triangular boundaries \cite{kumar2014}. We emphasize that nanodomain realignment and faceting is observable only because our experiments uniquely involve free-standing, edge-supported nanodisks \cite{kumar2014}, since substrate interactions would prevent such domain restructuring.
Our phase-field simulations indicate that this nanodoman realignment occurs between charged and uncharged states of the PZT nanodisk. When the disk is uncharged, the usual flux closure is observed implying that the ferroelastic domain walls are normal to the surface.  When it is charged, the polarization at the surface becomes radial and a vortex is formed at the center. In HRTEM, such a state is revealed by its nanodomains parallel to the surface. The faceting is selected to minimize both polarization and external anisotropies, principly due to strain fields at the perimeter due to the underlying (111) crystal axes of PZT.
This electrostatic analysis accounts for the two observed nanodomain configurations, but suggests a frequency response on gigahertz scales in contrast to the subhertz frequencies observed \cite{kumar2014}.  Because thermal effects have been ruled out for these experimental conditions \cite{zheng}, this slow time-scale suggests extreme nonlinearity and possibly mechanical overshoot. Our phase-field modelling indicates a driven and reversible charging/discharging cycle that is naturally described as an effective relaxation oscillation circuit \cite{vanderPol,Pippard79}. This is a system that is intrinsically out-of-equilibrium:  it cycles between an unstable state and a frustrated attempt to reach equilibrium that is prevented by onset behavior of its nonlinear element that forces it back to its original unstable state.  Electronically the nonlinear circuit element is often a neon light or a unijunction transitor, but its key feature is a threshhold voltage for negative resistance in parallel with a conducting channel that is usually a capacitor.

To our knowledge the first report of self-sustaining subhertz oscillations in a physical system was in a wide-gap semiconductor ZnO \cite{Losev25}, and we propose that the e-beam experiments on PZT nanodisks also present such an example. It is known that electron beam irradiation creates charge traps and defects in semiconducting targets.  The time-scale for the charging cycle will then be set by the charge saturation of these surface states. This charge storage threshhold in PZT targets exposed to electron beams has been measured quantitatively to be $5 \text{C/m}^2$. For the experiments of interest \cite{kumar2014} with current density of $0.5 \text{A/m}^2$, this then corresponds to a charging time of
\begin{equation}
\tau_{\text{charging}} = \frac{5 \text{C/m}^2}{0.5 \text{C/s-m}^2} = 10 \text{ sec}
\end{equation}
that is the same order of magnitude as the observed faceting oscillations \cite{kumar2014}.  We note that if we assume that these traps are singly charged, then there are $\frac{5 \text{C/m}^2}{1.6 \times 10^{-19} \text{C}/e} = 3 \times 10^{19} \rm{traps}/m^2$ that is consistent with concentrations reported for oxygen vacancies in commercial quality spin-on PZT films \cite{Mihara92}; this value is also in good agreement with the electron concentration used in our simulations ($N_e = 4 \times 10^{26} \text{m}^{-3}$) for 8 nm disks of thickness 100 nm.
Finally we note that our system is a charge analogue of a seesaw with a faucet continuously dripping water into a container on one end, and a block on the other one. In this textbook mechanical relaxational oscillator \cite{Wang1999}, the seesaw flips when the weight of the water container is greater than that of the block; however once the container touches the ground it empties, the seesaw goes back to its original position and the whole process starts again.  In the context of our present work, the electron beam is analogous to the water flow from the faucet; the threshold is related to the Zener breakdown \cite{Zener1948} that occurs in pn-junctions at low fields ca. $E \ll 1 \frac{\text{V}}{\text{nm}}$ and involves excitation of electrons to the conduction band from the valence band or trap states within the bandgap.  It is non-destructive and reversible, and thus differs from avalanche breakdown that  occurs at higher fields.
The charging of the nanocrystal creates depolarization fields that realign the domains and lead to the observed faceting (Fig 1).  During the initial part of the cycle, the current from the electron beam is filling the surface states with charge and the nanodisk behaves like a capacitor. However once the charge saturation threshhold for these traps and defects in the PZT disk is reached \cite{Li06}, the depolarization fields become strong enough to  initiate the current flow in the nanodisk. This current leads to discharging and the subsequent decrease of the depolarization field that caused the faceting; consequently the domains realign back to their uncharged configurations. The presence of surface states and large electromechanical coupling means that this involves domain creep in a random elastic medium that has been discussed quantitatively and measured in PZT films \cite{paruch}.  For the experimental parameters of interest, the domain wall creep in PZT epitaxial films is found to $v = 10^{-9}$ m/s \cite{paruch} so that for the dots of roughly 10 nm (of thickness 100 nm), the discharging time-scale is
\begin{equation}
\tau_{\text{discharging}} = \frac{10 \text{nm}} {10^{-9} \text{m/s}} = 10 \text{ sec}
\end{equation}
that is again in the same order of magnitude as the observed time-scale.  This charging/discharging cycle is expected to continue as long as the electron beam is on and a there is a continuous flow of charge into the nanodisk.  In contrast with many relaxation oscillators, in these PZT nanodisks there are slow time-scales associated with both the charging and the discharging cycles where the rate-limiting one is clearly the slower of the two.
Thus the cycling times, achieved both by macroscopic (charging threshhold) and microscopic (discharging via domain creep) methods, is consistent with measurement. We note here that the two underlying phenomena, charge trapping and slow domain creep due to electromechanical coupling and mass, are both absent in magnetism which explains the many orders of magnitudes that separate the frequency response of field-driven ferroelectric and ferromagnetic nanodisks.
In summary we have studied the mechanism underlying nanodomain restructuring in PZT nanodisks driven by electron beams.  Our phase-field modelling identifies the two observed nanodomain patterns with uncharged and charged free-standing boundaries; furthermore anisotropy in the charged case selects faceting at the perimeter to minimize electrostatic and elastic energy costs.  The observed subhertz frequency response, many orders of magnitude lower than expected from an electronic mechanism, is explained as relaxation oscillations where the charging/discharging times are determined by charge trap saturation and nanodomain wall creep. We predict that the oscillation time-scale can then be tuned by changing the surface trap density, either by varying the beam current or by annealing the PZT nanodisks in oxygen/ozone environments.  Though this charging/discharging cycle here is driven by electron beams, similar behavior could be achieved by gating the PZT nanodisks to achieve a voltage-controlled source-gate-drain device.  It would be unusual to have such small semiconductor triodes with such slow time-scales, and they could be important for biological applications like medical implants where time-scales are naturally of the order of seconds.

This work was supported by the A*STAR Computational Resource Centre through the use of its high performance computing facilities and by National Science Foundation grant NSF-DMR-1334428 (P. Chandra). PC is grateful for the hospitality of Trinity College, Cambridge where part of this work was performed.  We thank Robert Laskowski for useful discussions.
\bibliographystyle{ieeetr}
\bibliography{faceting7}
\end{document}